
\input phyzzx

\Pubnum {TIFR/TH/91--59}
\date {December 1991}
\titlepage
\title {THREE MANIFOLDS AND GRAPH INVARIANTS }
\author{S. KALYANA RAMA \foot{e-mail : kalyan@tifrvax.bitnet}
AND SIDDHARTHA SEN \foot{e-mail : sen@maths.tcd.ie }
\foot{Permanent Address : School of Mathematics, University
of Dublin, Trinity College, Dublin 2, IRELAND} }
\address {Tata Institute of Fundamental Research,\break
Homi Bhabha Road, Bombay 400 005, INDIA}
\abstract {
We show how the Turaev--Viro invariant can be understood within the
framework of Chern--Simons theory with gauge group SU(2). We also describe
a new invariant for certain class of graphs by interpreting the
triangulation of a manifold as a graph consisiting of crossings and
vertices with three lines. We further show, for $S^3$ and $RP^3$, that the
Turaev-Viro invariant is the square of the absolute value of
their respective partition functions
in SU(2) Chern--Simons theory and give a method of
evaluating the later in a closed form for lens spaces $L_{p,1}$.
}
\vfill


\endpage

\NPrefs
\def\define#1#2\par{\def#1{\Ref#1{#2}\edef#1{\noexpand\refmark{#1}}}}
\def\con#1#2\noc{\let\?=\Ref\let\<=\refmark\let\Ref=\REFS
         \let\refmark=\undefined#1\let\Ref=\REFSCON#2
         \let\Ref=\?\let\refmark=\<\refsend}

\define\TV
V. G. Turaev and O. Y. Viro, {\it State sum Invariants of 3--Manifolds
and Quantum 6--j Symbols}, LOMI Preprint (1990).

\define\RP
T. Regge, Nuovo Cemento {\bf 19} (1961) 558;
G. Ponzano and T. Regge, in Spectroscopic and Group Theoretic Methods
in Physics, Ed. F. Bloch (North--Holland, Amsterdam,1968).

\define\OS
H. Ooguri and N. Sasakura, Preprint KUNS 1088, HE(TH) RIMS-778 (August
1991).

\define\MT
S. Mizoguchi and T. Tada, Preprint YITP/U-91-43 (October, 1991).

\define\GV
M. Gross and S. Varsted, Preprint NBI-HE-91-33 (August 1991).

\define\D
U. Danielsson, Phys. Lett. {\bf B220} (1989) 137.

\define\FG
D. S. Freed and R. E. Gompf, Phys. Rev. Lett. {\bf 66} (1991) 1255.

\define\Ma
B. Hou, B. Hou and Z. Ma, Comm. Theor. Phys {\bf 13} (1990) 181.

\define\W
E. Witten, Comm. Math. Phys. {\bf 121} (1989) 351.

\define\Wi
E. Witten, Nucl. Phys. {\bf B322} (1989) 629;
{\it ibid} {\bf B330} (1990) 285.

\define\M
J. R. Munkres, Elements of Algebraic Topology, The Benjamin/Cummings
Publishing Company Inc., Menlo Park, CA, U.S.A, 1984.

\define\ST
H. Seifert and W. Threlfall, Seifert and Threlfall : A Text Book of
Topology, Academic Press, New York, NY, U.S.A, 1980.

\define\RB
R. Balasubramanian, private communication.

\define\KR
A. N. Kirillov and N. Yu. Reshetikhin, Adv. Series in Math. Phys. Vol. 7,
Ed. V. G. Kac, World Scientific (1988) 285--339.

Following the discovery of the Jones invariants in knot theory there has
been renewed interest in trying to find new invariants associated with
knots, graphs and 3--manifolds. A significant step was taken by Witten
when he was able to interpret the Jones invariants by using ideas from
quantum field theory \W\ .
Recently Turaev and Viro introduced \TV\ a new invariant
of $M_3$ ($M_d$ denotes a d--manifold). This invariant is combinatorial in
nature and is defined as a state sum computed on a triangulation of the
manifold and is based on the quantum 6--j symbols associated with the
quantised universal enveloping algebra, $U_q (SL(2,C))$. Here
$ q = exp(i { {2 \pi} \over {k+2} })  \, , k > 0 $ is a complex root of
unity. The Turaev--Viro (TV) invariant is also of interest for the
following reason. If one regards $j_i$'s of the classical 6--j symbol as
the lengths of the sides (colors) of a tetrahedron, then in the large
$j$ limit the positive frequency part of the 6--j symbol becomes
$e^{i S_R}$ where the Regge action $S_R$ is the discretised
version of the Euclidean
Einstein--Hilbert action $\int d^3 x \sqrt{g} R $ for 3--d gravity \RP\ .
However, the sum over the coloring of the tetrahedra is divergent in the
classical case but the quantum 6--j symbols provide a natural cutoff
$j \le {k \over 2}$ and regulate the divergent behaviour \OS\MT\ .
Thus the TV invariant is seen to be closely related to the partition
function of the 3-d gravity action.

A few remarks on the relation between the TV invariant $I_{TV}$ and other
known invariants of $M_3$ are also in order.
For orientable $M_3$ of the type
$M_2 \times [0,1] $ where $M_2$ is closed, the authors of \OS\ mention
that they have examined the partition functions of the TV
model for lower genus $M_2$ and found them to be equal to
the absolute--value--square of the partition function of the SU(2)
Chern--Simons theory. They also mention that the equality of these two
partition functions
has been proved by Turaev in a rather different approach.

In this paper we will show how the TV invariant can be understood within
the framework of Chern--Simons (CS) theory in which the gauge group is
SU(2). We evaluate the TV invariant for the manifolds $S^3$ and $RP^3$ and
show that for these manifolds $I_{TV} (M) = I_W^2 (M) $, where
$I_W (M) $ is the invariant for the manifold $M$ obtained by calculating
the absolute value of the partition function for a CS gauge theory with
gauge group SU(2). We will also attempt to understand the TV invariant in
terms of graphs and describe how one obtains a new invariant for certain
class of graphs using these ideas. In our discussions we will always
take $M_3$ to be a compact orientable 3--manifold with no boundary.

We start by assuming that a given $M_3$ has been triangulated and we
consider the graph $G$ associated with the triangulation. For example if
two tetrahedra $T_1, \, T_2$ are glued together along the face $F$ as
shown in figure 1 then we regard  $T_1, \, T_2$ as graphs rather than as
3--dimensional objects. The important observation we make regarding $G$ is
that $G$ is not an arbitrary graph; it is obtained  by gluing together a
collection of tetrahedra $\{ T_i \}$, represented symbolically as
$$G = \bigcup_{ \{ g_{ij} \} } \; T_i \eqn\one $$
where
$\{ g_{ij} \}$ encodes gluing information. Because of this if we want to
evaluate the invariant associated with $G$ by evaluating $G$ with respect
to a CS measure we have to specify what exactly the gluing process means
within this framework. First observe that a given tetrahedron can be
regarded as a collection of Wilson lines, joined together at each vertex by
an appropriate invariant coupling factors as discussed by Witten \Wi\ .
Each line may carry a different representation of SU(2). Gluing two
tetrahedra along face $F$ we will take to mean that the faces are
connected by ``walls" all carrying the trivial representation of SU(2) and
the representations on the common glued face are summed over. By repeatedly
using the factorisation technique of Witten \W\Wi\ ,
and observing that if a given
tetrahedron is enclosed in a ball with surface $S^2$ which intersects these
trivial representation--walls then the Hilbert space of $S^2$ is one
dimensional, it is easy to see that
$$ I(G) = \sum_{rep} \prod_i {\cal T}_i  \eqn\two $$
where $I(G)$ is the graph invariant and ${\cal T}_i$ is an object
associated with the tetrahedron
$T_i$ defined by its six sides each of which carries a
representation of SU(2). At this point we would like to point out the
factorised nature of the invariant $I(G) $ over the constituent tetrahedra
$T_i$. We will see shortly that there is a solution for ${\cal T}_i $, an
object associated with the tetrahedron $T_i$ which respects the
factorisable property.
However this solution will not have the required tetrahedral symmetry and
will lead to a modified prescription for the graph invariant.
Since $G$ does not have any boundary each ``face"
of a $T_i$ in $G$ is glued to some other face of a $T_j$ in $G$. Hence all
the representations appearing in  $\prod_i {\cal T}_i$
in equation \two\ are to be summed
over. Note that the graph $G$ is essentially the spine of a triangulated
manifold \TV\ and hence is inherently 3--dimensional. In particular the lines
in $G$ never cross and an arbitrary number $(\geq 3)$ of them can meet at
a vertex.

We will now attempt to understand the TV invariant in terms of planar
graphs with crossings and vertices with fixed number of lines,
and their invariants.
As before we consider the triangulation of $M_3$ as a set
of tetrahedra and ``gluing instructions".
The set of tetrahedra with a given coloring can be viewed as a
set of graphs and the gluing instructions a way of joining the graphs.
First, a tetrahedron with the colors $(a,b,c,d,e,f)$ labelling its sides
can be represented as a graph in any of the four ways shown in figure 2
where the crossings have a relative phase factor \KR\ .
In this graph each line and region is assigned a color. The colors of a
line and its neighbouring two regions as also the  colors of the three
lines joining at a vertex, which we call a 3--joint,
form an unordered triplet. Thus in
figure 2 the triplets are $(abc), \, (aef), \, (bdf) \, $ and $(cef)$. A
triplet $(abc)$ is said to be admissible if
$\vert a-b \vert \le c \le (a+b) $. In what
follows we will consider only those colorings with admissible triplets.

We implement the gluing instructions in this case as follows.
When two tetrahedra
with colors $(a,b,c,d,e,f)$ and $(a,b,c,l,m,n)$ are glued along their
common face with colors $(a,b,c)$, the graphs representing them are joined
together such  that the common area and the lines will have the same colors
as shown in figure 3. In this way the gluing instructions will lead to planar
graphs with crossings and 3--joints only.
Since the TV invariant does not have any phase factors to be associated
with crossings we will implement the gluing instructions in such a way
that in the resulting graph these phase factors will cancel out. This will
restrict us to a certain class of graphs only as described below.
It can be easily seen that representing the tetrahedra by crossings and
joints and gluing them may result in more than one
graph. It is also clear that the graph corresponding to a manifold without
boundary will be closed consisting only of closed lines.
The characterisation of the graphs obtained by the triangulations of a
given 3--manifold as described above is a difficult subject and is under
further study.

For future use we now define several quantities in a given graph. First we
define a quantity we call character for each given crossing.
It is a sum total of the colors of the four regions around the
crossing each with an appropriate sign which
is determined to be $+$ or $-$ according
to whether the region with that color comes to the right or left,
respectively, as one moves towards the crossing along the over--line. Thus
the characters of the crossings in figures 2B and 2C are $(b+e-c-f)$ and
$(c+f-b-e)$ respectively. Note that switching one crossing into another
(allowed by tetrahedral symmetry) reverses the sign of the character.
We call the sum of the characters of all the
crossings in a graph as the character of the graph, assigning 3--joints
a zero character.
The charcter of the graph is closely related to the sum total of the phase
factors associated with each crossings. In particular these phase factors
will cancel out whenever the character of the graph vanishes. Since the TV
invariant does not have any phase factors associated with the crossings it
is necessary for the graph to have a vanishing character.
Henceforth we will consider only
such graphs.

We also define $C_3$ and $C_4$ to be the
total number of 3--joints and crossings respectively; $L$ the total
number of lines; $l_i$ the total number of pieces one obtains by cutting a
given line $i$ at all crossings and 3--joints;
$\lambda$ the sum of $l_i$'s and $R$ the total number of regions.
When a given set of three lines all begin and end
in 3--joints we call them the ``Baryon
Orbits" (following \Wi\ ) and we define $B$ to be the
total number of baryon orbits. The
significance of these quantities is the following.  Let $N_0, \, N_1, \,
N_2 \,$ and $N_3$ denote the number of vertices, lines, faces and
tetrahedra respectively in the triangulation of the manifold
that gave rise to
the given graph. Then $N_3 = C_3 + C_4 , \; N_2 = C_3 + \lambda - B,
\; N_1 = L + R \;$ and  $N_0 = \chi + R + C_4 + L + B - \lambda$, where
$\chi$ is the Euler characteristic of the 3--manifold. Note that $\chi$
vanishes for $M_3$, a compact orientable 3--manifold with no boundary that
we are considering here.

Now given such a graph we would like to associate an object ${\cal T}$ for
each 3--joint and crossing (along with a phase factor \KR\ )
and define a quantity as in \two\ which will be
invariant under Reidemeister (R--)moves.
An invariant for which these properties are true is obtained by
identifying for each tetrahedron
$$ {\cal T} \equiv {\cal T}_W = \{ \;\;\; \}_q , \eqn\three $$
where $ \{ \;\;\; \}_q , $ denotes the quantum 6--j symbol. However this
object has a geometrical shortcoming. It does not have the symmetries of
a tetrahedron. There is an object related to   $ \{ \;\;\; \}_q  $ with
the symmetries of a tetrahedron which is none other than the quantum
Racah--Wigner coefficient upto some symmetry--preserving phase factors
and is given by
$$ {\cal T} \equiv {\cal T}_{TV} = (-1)^{-(a+b+c+d+e+f)}
\left\{\matrix{a & b & c \cr
          d & e & f}\right\}^{RW} \eqn\four $$
for each tetrahedron colored $(a,b,c,d,e,f)$ as in figure 2A.
But $I(G) $ in \two\  calculated using this object is not
invariant under R--moves. The problem which we consider is this : can we
modify the RHS of \two\ so that the tetrahedral symmetries are present and
it represents an invariant of the graph $G$ i.e. the R--moves leave
$I(G)$ invariant? This problem can be fixed,
while preserving the tetrahedral symmetry, by associating a factor
$ (-1)^{2 j} { {S_{0j}} \over {S_{00}} }$
with each line or region colored $j$ and by
requiring the graph to have vanishing character.
We then take the sum of this quantity over all possible coloring with
admissible triplets only. Thus the quantity
$$ I'_G = \sum_{coloring} \;\; \prod_{lines,regions} \;\; (-1)^{2 j}
{ {S_{0j}} \over {S_{00}} } \;\;
\prod_{crossings,3-joints} {\cal T} \eqn\five $$
satisfies our requirements of tetrahedral symmetry and is invariant under
Reidemeister moves. Notice, however, that in this process
the original factorisation property is lost.

In the above analysis we required tetrahedral symmetry for each crossing
and 3--joint because they represent actual tetrahedra in the triangulation
of the manifold which was the origin of our graph. Invariance under
Reidemeister moves are necessary because any graph is defined only upto
these moves.
However since the graphs we consider are obtained from (or can be
interpreted as) the triangulations of a 3--manifolds it follows that
$I'_G$ must be the same for any set of graphs corresponding to different
triangulations of the same 3--manifold. For the case of 3--manifolds it is
known that any two triangulations describe the same manifold if these
two are connected by a finite number of $(k,l)$ moves and their inverses
where $k+l = 5$ \GV\ . This set of moves is equivalent to Alexander moves
and to Matveev moves and bubble moves \TV\GV\ . The $(2,3)$
move corresponds to splitting two tetrahedra $OABC$ and $XABC$ into three
tetrahedra $OXAB, \; OXBC$ and $OXCA$ by joining the two vertices $O$ and
$X$. In terms of graphs this move corresponds to the Reidemeister--3 move.
The $(1,4)$ move corresponds
to obtaining four tetrahedra from one by adding a new vertex inside a
tetrahedron and connecting it to its four vertices. In terms of graphs
this corresponds to the ``bubble move" shown in figure 4.
(The inverses of the above moves are obvious).

Thus we require the
quantity $I'_G$ to be invariant not only under the Reidemeister moves but
also under the bubble moves.  However invariance under the bubble move
requires that we add a vertex dependent factor and hence, the right
candidate is found to be
$$ {\cal I}_G = S_{00}^{2 N_0} I'_G \eqn\six $$
where $N_0 = R + C_4 + L + B - \lambda$ is the total number of vertices
as defined before. Thus we obtain ${\cal I}_G$,
an invariant of a given graph with vanishing character.
We consider in the following the examples of $S^3$ and $RP^3$
in which we check
explicitly that two different triangulations lead to the same result.
That ${\cal I}_G$ is indeed independent of triangulation for any
manifold $M_3$  has been proved by Turaev and Viro \TV\ , which can also
be seen easily in our approach using the $(k,l)$ moves.

Thus we see that interpreting the triangulation of a
manifold as a planar
graph with crossings and vertices with fixed number of lines
one obtains a new invariant which is defined for
all such graphs with vanishing character. This quantity remains invariant
under Reidemeister moves and bubble moves
for the graphs and gives TV invariant when the
graph is viewed as a triangulation of a manifold. This suggests that any
graph with vanishing character can be interpreted as a triangulation of
3--manifold and has ${\cal I}_G$ as its invariant.
(However, as mentioned earlier, the full classification of graphs which
correspond to the triangulation of a 3--manifold is a difficult subject
and is under further study). As can be seen
from such an interpretation ${\cal I}_G$ is invariant not
only under Reidemeister moves and bubble moves
but also under `cutting off the crossings and
3--joints and interconverting them and gluing  them   back preserving the
``gluing instructions" and the character of the graph' -- such a process will
in general lead to a different graph. This process of reducing by symmetry
transformations the original graph to a set of standard simple
objects whose invariants can be evaluated easily is quite akin to the
skein relations in calculating the usual graph invariants.
To evaluate ${\cal I}_G$ completely we need one further identinty
$$ \sum_{a,b} \;\; b \;\;\; \Biggl\vert a \;\;\; c \;\;\;\;
{ {S_{0a}} \over {S_{00}} } \; { {S_{0b}} \over {S_{00}} } \;\; = \;\;
{ 1 \over {S_{00}^2} } \; { {S_{0c}} \over {S_{00}} } \eqn\aone  $$
where $a, \; b$ and $c$ label the line and its neighbouring regions
respectively.

We illustrate the above procedures for a simple case of two tetrahedra
glued together as in figure 5a. (Actually, this figure represents
one way of triangulating $S^3$). This graph has a vanishing character.
Replacing the two tetrahedra (3--joints) by two different crossings we
obtain the graph of figure 5b which has a non vanishing character and
hence is not admissible. Switching one of the crossings lead to an
admissible graph with vanishing character (figure 5c) which, under a
Reidemeister move (same as summing over $b_1$), leads to the graph in
figure 5d. Note that one could have obtained the graph in figure 5d from
that in figure 5a in a single step by using the Reidemeister move
for the 3--joint. Thus this example, as well as all the others that we have
tried, indicates that irrespective of how one initially represents the
given set of tetrahedra as a graph preserving the gluing
instructions one always gets the same result for ${\cal I}_G$ after
a sufficient number of symmetry operations -- as one must if ${\cal I}_G$
were a genuine invariant of a graph with vanishing character
(or equivalently of a triangulated manifold).

We now evaluate the TV invariant for $S^3$ and $RP^3$. The 3--sphere $S^3$
can be viewed as the boundary of a 4--simplex with the vertices denoted by
$0,1,2,3$ and $4$. Thus $S^3$ consists of five tetrahedra whose vertices
are $(0123), (0124), (0134), (0234) $ and $(1234)$. The graph
corresponding to this triangulation is given in figure 6. The graph
invariant ${\cal I}_{S^3}$ (which is the same as TV invariant) for $S^3$
can be evaluated easily using the Reidemeister moves and the result is
$${\cal I}_{S^3} = S_{00}^2 .  \eqn\seven  $$

The 3--projective space $RP^3$ can be triangulated as shown in figure 7
where the lines labelled the same, and the corresponding faces, are to be
identified. The TV invariant for $RP^3$ can be calculated easily and is
given by
$${\cal I}_{RP^3} = { 2 \over {k+2} } (1 + (-1)^k)
{\rm sin}^2  { \pi \over { 2 (k+2) } }  \eqn\eight $$
where $k$ is related to $q$, the root of unity, as defined earlier.

One can obtain another triangulation of $S^3$ and $RP^3$ by first
triangulating a lens space $L_{p,q}$ since the manifolds $S^3$ and $RP^3$
are special cases of $L_{p,q}$ with $(p,q) = (1,0) $ and $(2,1)$
respectively. The lens space $L_{p,q}$ is obtained as
follows \ST\ . Consider a region of 3--space bounded by two spherical caps
meeting in an equatorial circle. Rotate the lower cap onto itself through
an angle of ${ {2 \pi q} \over p }$ radians and then reflect it about the
equatorial plane onto the upper cap. The resulting manifold thus obtained
is the lens space $L_{p,q}$. One possible triangulation $L_{p,q}$,
which suffices for our purposes, is shown in figure 8 with the lines
labelled the same and the corresponding faces identified \M\ . In figure 8,
$ i \in {\bf Z} / p $ where $i$ is the subscript of the   label $\beta_i$.
Using this triangulation the expression for the TV invariant
${\cal I}_{L_{p,q}} $ can be written in terms of ${\cal T} $ for each
tetrahedron. For $(p,q) = (1,0) $ and $(2,1)$, this expression can be
evaluated and the results agree with equations
\seven\ and \eight\ ,
as they should. However, this agreement would not be there if the vertex
factor in equation \six\ were absent.  Though we find it hard to
evaluate the expression for ${\cal I}_{L_{p,q}} $ in general, we are
able to evaluate it for
another case, $(p,q) = (3,1) $. But in this particular instance,
with the triangulation for $L_{3,1} $ given as in figure 8, the proof for
the invariance of ${\cal I}_{L_{p,q}} $ is not valid for reasons given in
\TV\ and hence ${\cal I}_{L_{3,1}} $ evaluated as above is not the right
answer.

In a different context Danielsson \D\ has evaluated $I_W (M)$ for
$M = S^3$ and $L_{p,1}$, where
$I_W (M) $ is the invariant for the manifold $M$ obtained by calculating
the absolute value of the partition function for a CS gauge theory with
gauge group SU(2). They can be written as
$$\eqalign{
I_W (S^3) & = \sqrt{ {2 \over {k+2}} } {\rm sin} {\pi \over {k+2}}  \cr
I_W (L_{p,1}) & = {2 \over {k+2}}  \;\; \Biggl\vert \; \sum_{n=0}^{k+1}
{\rm sin}^2 {{\pi  n} \over {k+2}}
exp ( i {{\pi  p n^2} \over {2 (k+2)}} ) \; \Biggr\vert . } \eqn\nine $$
Comparing our results for ${\cal I}_{S^3}$ and ${\cal I}_{RP^3}$
with equation \nine\ after setting $p=2$  (see below for evaluation of
$I_W (L_{p,1})$) we find that
$$ {\cal I}_M = I_W^2(M) \eqn\ten $$
for $M = S^3$ and $RP^3$.

The sum in equation \nine\ , denoted by
${ 1  \over 4 } \sigma_{p,r} , \; r = k+2 $
can be evaluated as follows. (This procedure is due to R. Balasubramanian
\RB\ .)  First, it can be seen that
$$ \sigma_{p,r} = { 1 \over 2 } ( G(p,0,4r) - G(p,4,4r) )  \eqn\eleven $$
where
$G(a,b,l) = \sum_{n=0}^{l-1} exp ( i { {2 \pi} \over l }
( a n^2 + b n) ) $.
Now we state the following properties of $G(a,b,l)$ \RB\ .
In the following all the variables are integer valued and  $(x,y)$ denotes
the greatest common divisor of two integers $x$ and $y$.

(1) If $(a,l)$ does not divide $(b,l)$ then $G(a,b,l) = 0$.

(2) $G(ca,cb,cl) = c G(a,b,l) $.

Using the properties (1) and  (2) we
will restrict ourselves to the case where $(a,l) = 1$.

(3) Let $\omega$ be an integer such that $ 2 a \omega + b $ is a multiple
of $l$. Such an $\omega$ is guaranteed to exist for any $a,b$ and $l$
provided $(2a,l)$ divides $(b,l)$. Since $(a,l) = 1$, this implies that
$\omega$ exists if either $l$ is odd or if both $b$ and $l$ are even. Then
$ G(a,b,l) = exp ( - i { {2 \pi}  \over l } a \omega^2 ) G(a,0,l)$.

(4) If $l$ is even and $b$ is odd then
$ G(a,b,l) =  exp ( - i { {2 \pi}  \over l } a \mu^2 ) G(a,0,l)$
where  $\mu$ is an integer such that
$ 2 a \mu + (b - 1) $ is a multiple of $l$.

(5) For $l$ odd $\vert G(a,0,l) \vert = \sqrt{l} .$

(6) Let $l = 2 \lambda$. Then
$\vert G(a,0,l) \vert^2 + \vert G(a,1,l) \vert^2 = 2 l $
and $\vert G(a,0,l) \vert = 0 $ (or $ \sqrt{2 l} $) if
$a \lambda$ is odd (or even).

Using the above properties (1) -- (6) of $G(a,b,l), \;  I_W (L_{p,1}) $
in \nine\ can be evaluated for any $p$ and $r$. For example, one can
obtain all the values of  $I_W (L_{p,1}) $ listed in \D\ for some specific
values of $p$ and $k$.

Moreover, assuming that equation \ten\ is true for any
manifold ( see \OS\ ), the TV invariant for the lens space $L_{p,1}$ can
also be obtained by the above method.

\vskip 1.0cm

\centerline{ACKNOWLEDGEMENTS}

It is a pleasure to thank Sumit Das for many discussions and for bringing
the references \TV\OS\ to our attention. We would also like to thank
R. Balasubramanian for giving us his result on the summation in  equation
\nine\ .

\refout

\vfill

\endpage

\centerline{FIGURE CAPTIONS}

Figure 1 : Gluing procedure.

Figure 2 : Representation of a tetrahedron by graphs.

Figure 3 : Gluing procedure for graphs.

Figure 4 : Bubble move.

Figure 5 :  One example of graph manipulations.

Figure 6 : Graph of $S^3$.

Figure 7 : Triangulation of $RP^3$.

Figure 8 : Triangulation of $L_{p,q}$
(subscripts for $ \beta \in {\bf Z}/p $).

\end